\documentclass[fleqn,usenatbib]{mnras}

\usepackage{newtxtext,newtxmath}
\usepackage[T1]{fontenc}
\usepackage{ae,aecompl}

\usepackage{graphicx}	% Including figure files
\usepackage{amsmath}	% Advanced maths commands
\usepackage{amssymb}	% Extra maths symbols

\newcommand{\fref}[1]{Fig.~\ref{#1}}
\newcommand{\cms}{\ \ensuremath{{\rm cm}}\,\ensuremath{{\rm s}^{-1}}}
\newcommand{\ms}{\ \ensuremath{{\rm m}}\,\ensuremath{{\rm s}^{-1}}}

\newcommand{\avg}[1]{\langle #1 \rangle}

\newcommand{\vb}{\ensuremath{v_{\rm B}}}
\newcommand{\harpsn}{\mbox{HARPS-N}}

\title[Photon-weighted barycentric correction]{Photon-weighted barycentric correction and its importance for precise radial velocities}
\author[R. Tronsgaard et al.]{Ren\'{e} Tronsgaard,$^{1}$\thanks{E-mail: rtr@space.dtu.dk (RT)}  % ORCID 0000-0003-1001-0707
Lars A. Buchhave,$^{1}$     % ORCID 0000-0003-1605-5666
Jason T. Wright$,^{2}$      % ORCID 0000-0001-6160-5888
Jason D. Eastman,$^{3}$     % ORCID 0000-0003-3773-5142
\newauthor
and Ryan T. Blackman$^{4}$  % ORCID 0000-0002-0303-3276
\\
$^{1}$DTU Space, National Space Institute, Technical University of Denmark, Elektrovej 328, DK-2800 Kgs.
Lyngby, Denmark\\
$^{2}$Center for Exoplanets and Habitable Worlds, Department of Astronomy \& Astrophysics, The Pennsylvania State University, \\
525 Davey Laboratory, University Park, PA 16802, USA\\
$^{3}$Center for Astrophysics \textbar \ Harvard \& Smithsonian, 60 Garden St, Cambridge, MA 02138, USA\\
$^{4}$Department of Astronomy, Yale University, 52 Hillhouse Avenue, New Haven, CT 06511, USA
}

\date{Accepted 2019 July 25. Received 2019 July 9; in original form 2019 July 9.}

\pubyear{2019}

\begin{document}
\label{firstpage}
\pagerange{\pageref{firstpage}--\pageref{lastpage}}
\maketitle

% Abstract of the paper
\begin{abstract}
%The abstract should briefly describe the aims, methods, and main results of the paper.
%It should be a single paragraph not more than 250 words (200 words for Letters).
%No references should appear in the abstract.
When applying the barycentric correction to a precise radial velocity measurement, it is common practice to calculate its value only at the photon-weighted midpoint time of the observation instead of integrating over the entire exposure. However, since the barycentric correction does not change linearly with time, this leads to systematic errors in the derived radial velocities. The typical magnitude of this second-order effect is of order 10\cms{}, but it depends on several parameters, e.g. the latitude of the observatory, the position of the target on the sky, and the exposure time. 
We show that there are realistic observing scenarios, where the errors can amount to more than 1\ms{}. We therefore recommend that instruments operating in this regime always record and store the exposure meter flux curve (or a similar measure) to be used as photon-weights for the barycentric correction.
In existing data, if the flux curve is no longer available, we argue that second-order errors in the barycentric correction can be mitigated by adding a correction term assuming constant flux.
\end{abstract}

% Select between one and six entries from the list of approved keywords.
% Don't make up new ones.

%%%%%%%%%%%%%%%%%%%%%%%%%%%%%%%%%%%%%%%%%%%%%%%%%%

%%%%%%%%%%%%%%%%% BODY OF PAPER %%%%%%%%%%%%%%%%%%

\section{Introduction}
\label{sec:intro}
%All papers should start with an Introduction section, which sets the work
%in context, cites relevant earlier studies in the field by \citet{Others2013},
%and describes the problem the authors aim to solve \citep[e.g.][]{Author2012}.
\noindent In order to search for and characterise potential Earth-analogues outside our own Solar System, astronomers will need precise radial velocity (PRV) instruments that can deliver a yet unseen precision of a few\,\cms{} over timescales of several years \citep[e.g.][]{Fischer:2016}. Among the numerous prerequisites for being able to make such measurements is the ability to accurately transform the measured Doppler shift into a stationary reference frame with respect to the barycentre of our Solar System. 

\citet{Wright:2014} presented a detailed review of how to calculate and perform  barycentric corrections to an accuracy better than $~1\cms$. The authors provided an IDL implementation \texttt{ZBARYCORR}, which has later been ported to Python \citep{Kanodia:2018}. However, one important question that has not been fully addressed in literature is how to best apply the instantaneous barycentric correction to an exposure that is extended in time. During the integration time of an exposure, usually tens of minutes, there can be a significant change in the barycentric correction, mainly due to Earth's rotation, and the recorded spectrum will get slightly smeared out on the detector, effectively broadening the stellar absorption lines. Time-variable changes in the flux that reaches the spectrograph, caused by clouds, variable seeing, poor telescope guiding, or other issues, will make this smearing non-uniform, shift the weight of the broadened lines, and introduce spurious Doppler shifts. This type of error, which we will refer to in this paper as the first-order effect, is usually corrected for by monitoring the photon flux through the spectrograph with an exposure meter. From these measurements, the {\it photon-weighted  midpoint time} of the exposure can be determined, ensuring that the barycentric velocity is calculated at the mean time of arrival of the collected photons instead of the ``geometric'' midpoint time of the exposure, equidistant from the shutter open and close times. However, this method neglects the fact that the barycentric correction does not change linearly. As we will show in the following, a more correct approach is to weight the corrected velocity with the exposure meter flux curve and in that way capture the higher-order variation of the barycentric motion of the observatory. Neglecting what we will refer to as the second-order effect can lead to typical systematic errors of order $10\cms$ for exposure times of 30-60 minutes. In some quite realistic observing scenarios it will cause errors greater than $1\ms$.

The problem was first pointed out in \citet{Fischer:2016}, referring to an early draft version of this manuscript, and it was briefly mentioned again in \citet{Blackman:2017}, along with a description of the chromatic dependence of the barycentric correction. This paper aims to provide a more detailed and formal description of how to compute the barycentric correction for an extended exposure, estimate the error that arises when the second-order effect is neglected, and discuss some recommendations for future observations and existing data.

In Section~\ref{sec:correctbary}, we describe how to calculate the barycentric correction for an extended exposure with the best possible accuracy, and in Section~\ref{sec:derivation} we estimate the errors that result from using the geometric and photon-weighted midpoint times. 
In Section~\ref{sec:simulations}, we simulate the error using the full barycentric correction algorithm to confirm our analytic estimate and to understand how the error depends on the shape of the exposure meter flux curve. 
In Section~\ref{sec:discussion}, we discuss implications for exoplanet mass measurements, how to optimise observing strategies, and how to deal with and possibly improve existing data.

\section{Photon-weighted barycentric correction}
\label{sec:correctbary}
\begin{figure*}
    
    \begin{minipage}{0.49\textwidth}
        \includegraphics[width=\linewidth]{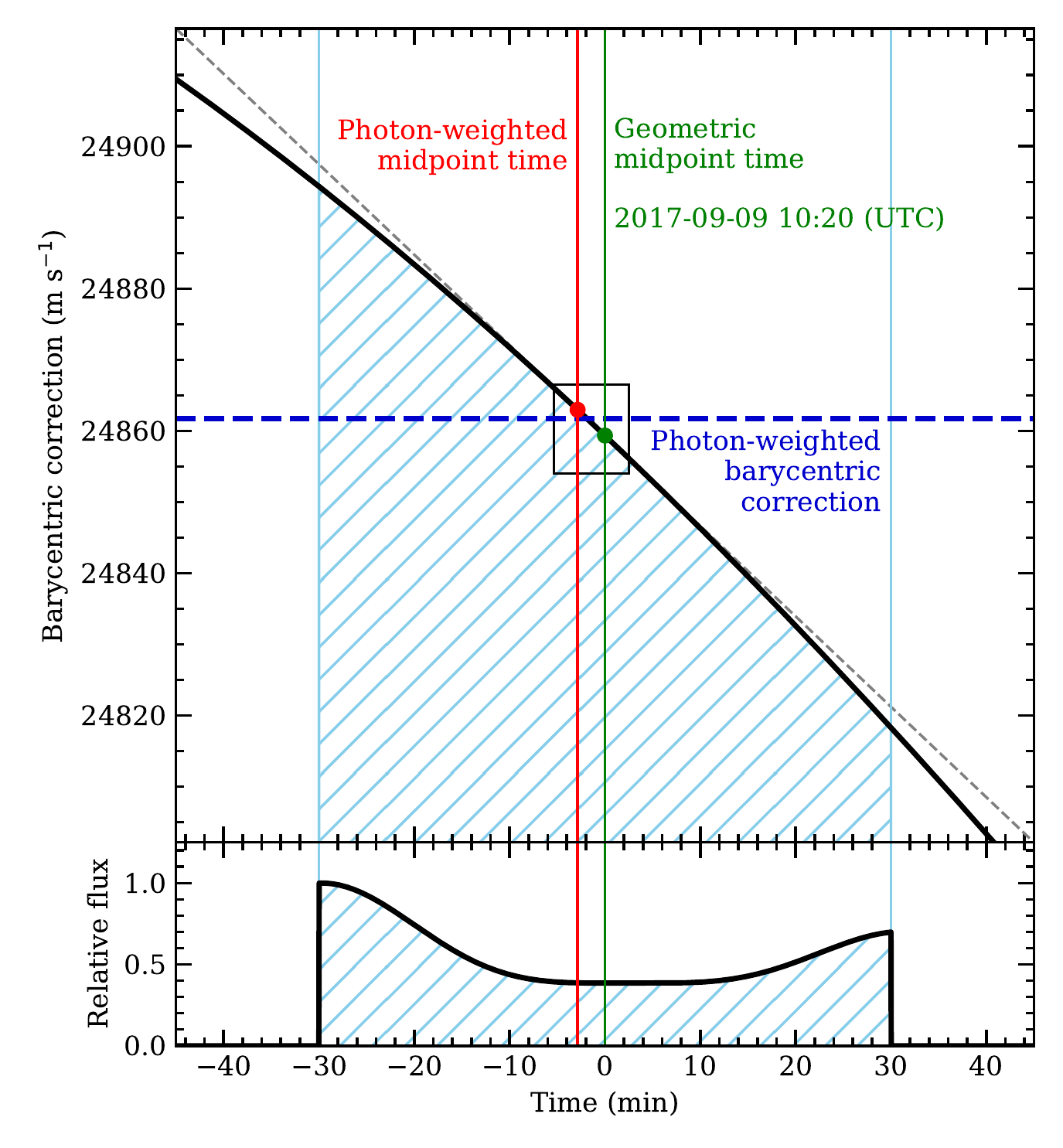}
    \end{minipage}
    \begin{minipage}{0.49\textwidth}
        \includegraphics[width=\linewidth]{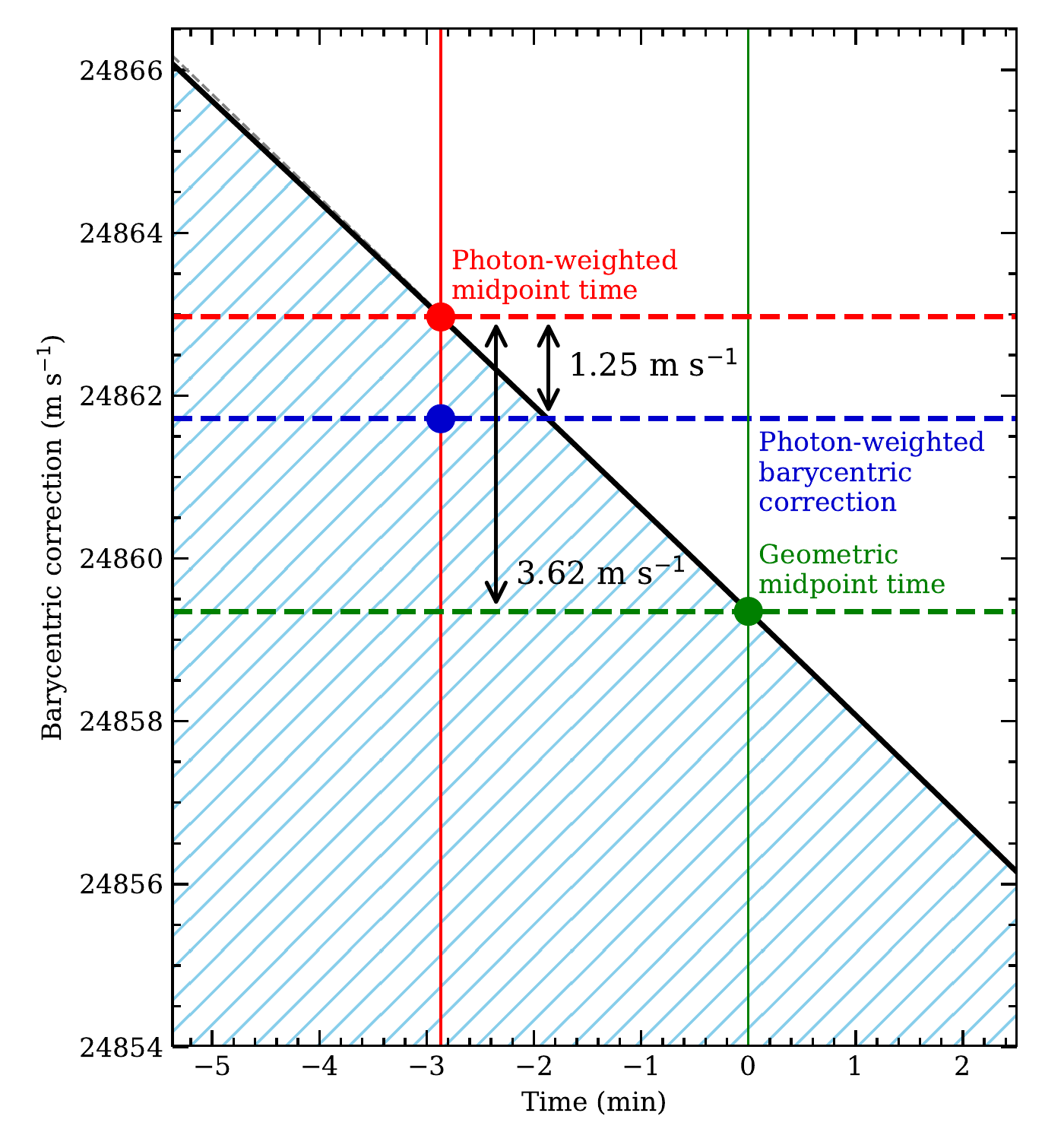}
    \end{minipage}
    \caption{Left panel shows the instantaneous barycentric correction (black curve) for a star during one 60-minutes exposure at Mauna Kea. The fictional, rising target is chosen such that the exposure starts when the star is $30^\circ$ above the horizon (air mass $=2.0$), and such that we are looking due east at the geometric midpoint time of the exposure. The lower panel shows what the exposure meter flux could look like during the exposure, e.g. if a thin cloud passes over the telescope or if the seeing changes. We use the flux curve to compute the photon-weighted midpoint time (red) and the photon-weighted barycentric correction (blue). The zero point of the time axis is set at the geometric midpoint time of the exposure (green). The dashed grey line is the tangent to the barycentric correction curve at the geometric midpoint time, emphasising the curvature of the barycentric correction function. The right panel shows an enlarged view around the centre of the exposure. In this particular case there is a difference of $1.25\ms$ between the photon-weighted barycentric correction (blue) and the barycentric correction at the photon-weighted midpoint time (red).}
    \label{fig:example}
\end{figure*}
\noindent Using the methods of \citet{Wright:2014}, we are easily able to calculate the barycentric correction $z_{\rm B}$ with a precision better than 1\cms{} for a photon arriving from a star to a telescope on Earth at a given point in time. As emphasised in that paper, $z_{\rm B}$ is a redshift, and it must be applied multiplicatively to the measurement, $z_{\rm meas}$, i.e.
\begin{align}
    1 + z = (1 + z_{\rm meas}) (1 + z_{\rm B}),
\end{align}
where $z$ is the ``true'' barycentric-motion-corrected redshift. Thus the corrected radial velocity is calculated as 
\begin{align}
    v = cz 
    = c \left[ \left(1 + \frac{v_{\rm meas}}{c}\right) \left(1 + \frac{v_{\rm B}}{c}\right) - 1 \right],
    \label{eq:v_true}
\end{align}
where we express the barycentric correction as a velocity, $v_{\rm B} = cz_{\rm B}$.

When taking an exposure of length $\Delta t$ in a spectrograph, one would ideally calculate the instantaneous barycentric correction for each individual photon at the time it arrives to the telescope, then average all the calculated values for the collected photons in each pixel on the detector. If $f(t)$ is proportional to the flux of photons in one pixel, and $v(t)$ denotes the barycentric corrected velocity as a function of time, the average corrected velocity of the exposure in that pixel can be expressed as
\begin{align}
    \langle v \rangle 
    = \left. \int_{t_0-\Delta t/2}^{t_0+\Delta t/2} v(t) f(t) dt 
      \middle/ \int_{t_0-\Delta t/2}^{t_0+\Delta t/2} f(t) dt \right.,
    \label{eq:v_avg}
\end{align}
where $t_0$ is the geometric midpoint time of the exposure. 

Depending on the type of detector, $f(t)$ can be obtained in various ways. When infrared arrays are used with ``up-the-ramp'' sampling, which means sampling the voltage of each pixel every few ($\sim$10) seconds throughout the exposure, it produces a direct measure of the time history of the photon arrivals. This is how instruments like the \textit{Habitable-Zone Planet Finder} \citep{Mahadevan:2012} are operated.
CCD detectors work differently, and it is not possible to directly monitor the flux received by the detector without reading out the image.
Instead one can use the exposure meter flux as a proxy. \citet{Landoni:2014} and \citet{Blackman:2017} describe how to use a multi-channel exposure meter to measure $f(t)$ for a particular wavelength associated with a pixel or a section of the detector.

In this paper, we will refer to $f(t)$ as the exposure meter flux curve, although our findings also apply to instruments with infrared arrays or other ways of monitoring flux variations. We will limit ourselves to a single wavelength channel (or a single pixel) and instead focus on quantifying the errors that can result from calculating the barycentric correction at a defined midpoint time rather than using Equation~\eqref{eq:v_avg}.
The difference between the geometric and photon-weighted midpoint times, and how the photon-weighted barycentric correction can differ from what is calculated at the midpoint times, is illustrated by an example in \fref{fig:example}.

\section{Quantifying the errors}
\label{sec:derivation}
\begin{figure*}
    \centering
    \includegraphics[width=\textwidth]{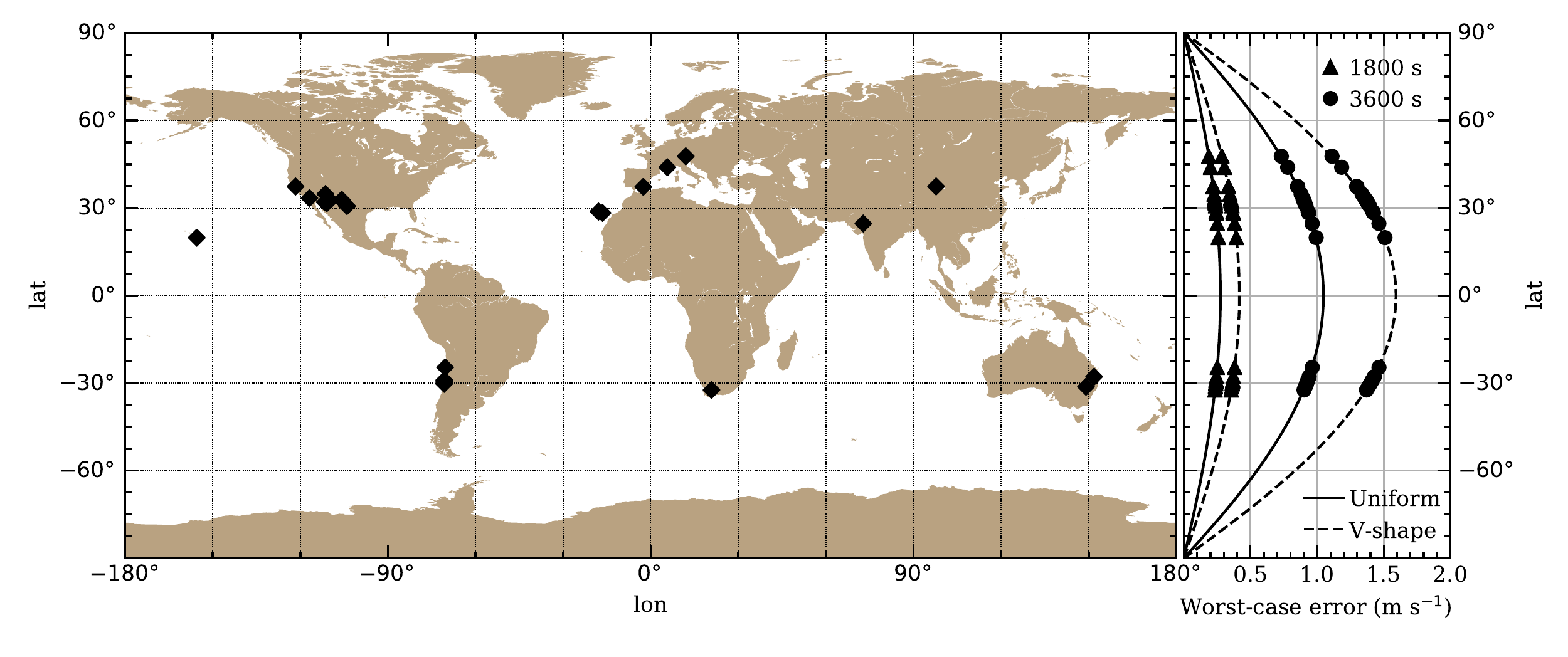}
    \caption{\textit{Left panel}: Locations of 25 different observatories that are hosting or will soon be hosting one or more PRV instruments \citep[see e.g.][]{Plavchan:2015,Fischer:2016,Wright:2017}. Most of these sites are located around latitude $\pm 30^\circ$, yielding $\cos({\rm lat})\approx 0.87$. Of these instruments, Mauna Kea (Hawaii) at latitude $+20^\circ$ is the PRV site closest to equator, with $\cos({\rm lat})=0.94$. 
    \textit{Right panel:} Systematic error for the realistic worst-case observing scenario. A rising (or setting) target due East (or West) is observed with the exposure starting (or ending) at $30^\circ$ altitude (air mass $=2.0$). Solid and dotted lines indicate the shape of the exposure meter flux curve. Markers indicate the exposure time.}
    \label{fig:worldmap}
\end{figure*}
\noindent On short timescales (hours), the change in the barycentric correction $v_{\rm B}$ is dominated by the diurnal rotation of the Earth. For a telescope at location $\rm (lat, lon)$ observing a star at position $(\alpha, \delta)$, the projected velocity of the telescope towards the star is given by
\begin{align}
    -\vb(t) 
        \approx \frac{2\pi R_\oplus}{24 {\rm h}} \cos({\rm lat}) \cos (\delta) \sin \left( \psi(t) \right) + {\rm constant}
    \label{eq:diurnal}
\end{align}
where $\psi(t)$ is the local hour angle, expressed in radians, written here in terms of the local sidereal time (LST):
\begin{align}
    \psi(t) = \frac{2\pi}{24 {\rm h}} {\rm LST}(t, {\rm lon}) - \alpha.
\end{align}
The hour angle, and thus $\vb(t)$, is a sinusoidal variation with a period of $23.93$ hours and amplitude  
\begin{align*}
    V_0 = \frac{2\pi R_\oplus}{24 {\rm h}} \cos({\rm lat}) \cos (\delta),
\end{align*}
which can be up to {463\ms} (equatorial velocity of the Earth). The time derivative of $\psi(t)$, which we will soon need, can be approximated as $2\pi/24{\rm h}$. For now, we consider a star with $v_{\rm meas}=0$, allowing us to set $v = \vb$

The photon-weighted midpoint time of an exposure, $\avg{t}$, can be defined in the same way as how we defined the average corrected velocity in Equation~\eqref{eq:v_avg}:
\begin{align}
    \langle t \rangle 
    = \left. \int_{t_0-\Delta t/2}^{t_0+\Delta t/2} t f(t) dt 
      \middle/ \int_{t_0-\Delta t/2}^{t_0+\Delta t/2} f(t) dt \right. .
    \label{eq:t_avg}
\end{align}
We can now expand $v$ around $\avg{t}$ with a second-order Taylor polynomial:
\begin{align}
    v(t') \approx v(\avg{t})
        - & 2 \pi V_0 \cos(\psi(\avg{t})) \frac{t' - \avg{t}}{24{\rm h}} \notag \\
        + & 2 \pi^2 V_0 \sin(\psi(\avg{t})) \left[\frac{t' - \avg{t}}{24{\rm h}}\right]^2.
    \label{eq:taylor}
\end{align}
If we define the first-order error as the difference between calculating the barycentric correction at the geometric and photon-weighted midpoints, respectively, it can be approximated as:

\begin{align}
    v(t_0) - v(\avg{t})
    & \approx -2.0\ms  \cos({\rm lat}) \cos(\delta) \cos(\psi(\avg{t})) \frac{t_0 - \avg{t}}{1~{\rm min}}
    \label{eq:firstorder}
\end{align}
A systematic error of up to 2\ms{} for every minute of difference between the geometric and photon-weighted midpoint times is a quite severe error in PRV context. It becomes largest when observing stars on the meridian, which is where most observers usually try to observe their targets. The community is well-aware of this effect, and it has become a standard procedure at almost every PRV instrument to correct for this by recording the photon-weighted midpoint time for each exposure.

By inserting Equation~\eqref{eq:taylor} into Equation~\eqref{eq:v_avg}, we can estimate the second-order error, which we define as the difference between the corrected velocity at the photon-weighted midpoint time, $v(\avg{t})$, and the photon-weighted average of the corrected velocity, $\avg{v}$:
\begin{align}
    v(\avg{t}) - \avg{v} \approx -2 \pi^2 V_0 \sin(\psi(\avg{t})) \frac{\avg{t^2} - \avg{t}^2}{(24{\rm h})^2}.
    \label{eq:taylor_weighted}
\end{align}
The second-order error is caused by the curvature of $\vb$, as shown in \fref{fig:example}. In principle, we are moving a smoothing filter over the instantaneous barycentric correction function, using the exposure meter flux curve as the smoothing kernel. This means that the point $(\avg{t},\avg{v})$ will always land on the inside of the curvature, and the second-order error will thus always have the same sign as the diurnal component of $v_{\rm B}$ (positive east of the meridian and negative west of the meridian). We also note that the magnitude of the second-order error will usually have a dependence on $\Delta t^2$.

Whereas the first-order error reaches its maximum when the target passes over the meridian, the second-order error gets largest at hour angle $\pm 6^{\rm h}$ and declination zero. This is where the celestial equator intersects with the horizon, and it is thus not observable. A more useful number is the maximum error possible above a certain altitude on the sky, and we therefore transform Equation~\eqref{eq:taylor_weighted} to horizontal (alt,az) coordinates. With a bit of spherical trigonometry\footnote{See for example \url{https://web.archive.org/web/20181013052114/http://star-www.st-and.ac.uk/~fv/webnotes/chapter7.htm}}, one can show that
\begin{align*}
    \cos(\delta)\sin(\psi) = - \cos({\rm alt}) \sin({\rm az}).
\end{align*}
If we consider a uniform exposure (i.e. $f(t)={\rm constant}$), we can evaluate the integrals contained in $\avg{t}$ and $\avg{t^2}$:
\begin{align}
    v(\avg{t}) - \avg{v}
        \approx -1.32 \ms \cos({\rm lat}) & \cos(\delta) \sin(\psi) \left( \dfrac{\Delta t}{1 \rm h} \right)^2 \notag\\
        \approx 1.32 \ms \cos({\rm lat}) & \cos({\rm alt}) \sin({\rm az}) \left( \dfrac{\Delta t}{1 \rm h} \right)^2 
    \label{eq:err_uniform}
\end{align}
The error is largest at ${\rm az}=\pm 90^\circ$, i.e. when facing straight east or west. We adapt a lower limit on the altitude of $30^\circ$ (air mass $=2.0$) based on the common practice we have seen at various PRV instruments. Since we would like to observe the entire exposure above the altitude limit, the actual limit for the `alt' parameter in Equation~\eqref{eq:err_uniform} depends on the latitude and exposure time. Therefore, in the following example for Mauna Kea (Hawaii), we set the altitude limit to $37^\circ$ for a 60~min exposure and $33.5^\circ$ for a 30~min exposure.

As shown in \fref{fig:worldmap}, the majority of current and planned PRV instruments are located at observatories near latitude $\pm 30^\circ$, with Mauna Kea at latitude $+20^\circ$ being the closest to equator. If we expose for 30~min at this latitude and employ worst-case altitude and azimuth angles, as described above, the second-order error becomes $0.25\ms$; if we double the exposure time to 60~min, the error quadruples to $1.0\,\ms$. In other words, using the photon-weighted midpoint time for the barycentric correction can lead to significant systematic errors even at the 1\ms{} RV precision level. 

For non-uniform flux curves, the result in Equation~\eqref{eq:err_uniform} needs to be altered only by a scaling factor. As an example, it can be shown that a V-shaped flux curve, symmetrically dropping to zero at the geometric midpoint time, yields an effect of exactly 1.5 times the uniform flux result. In general, if photons are missing in the middle of the exposure (e.g. due to a cloud passing over the telescope), the second-order error increases. If photons are concentrated towards one end of the exposure, the error second-order error decreases.

We finally note that although we obtained these results for $v_{\rm meas}=0$ in Equation~\eqref{eq:v_true}, they can safely be generalised to any $v_{\rm meas} \ll c$. The factor $(1+v_{\rm meas}/c)$ has a significant effect on the absolute value of the barycentric correction \citep{Wright:2014},
but its influence on the size of the second-order error is negligible even at the 1\cms{} precision level.

\section{Simulated observations}
\label{sec:simulations}
\noindent We have used the \texttt{barycorrpy} package \citep{Kanodia:2018} to simulate, under various conditions, the systematic second-order error that would result from not photon-weighting the barycentric correction. 

For each set of simulations, we let the same observation happen at the same time on multiple sky positions. We generate an exposure meter flux curve with a given shape, exposure time, and sampling cadence. Then, for a range of sky positions, we calculate the barycentric correction both as the photon-weighted average, $\avg{v}$, and at the photon-weighted midpoint time, $v(\avg{t})$. The integrals are carried out as simple Riemann sums, evaluated at the geometric midpoint of every exposure meter step. Finally, the second-order error is calculated as $v(\avg{t})-\avg{v}$. 

\begin{figure}
    \centering
    \includegraphics[width=\linewidth]{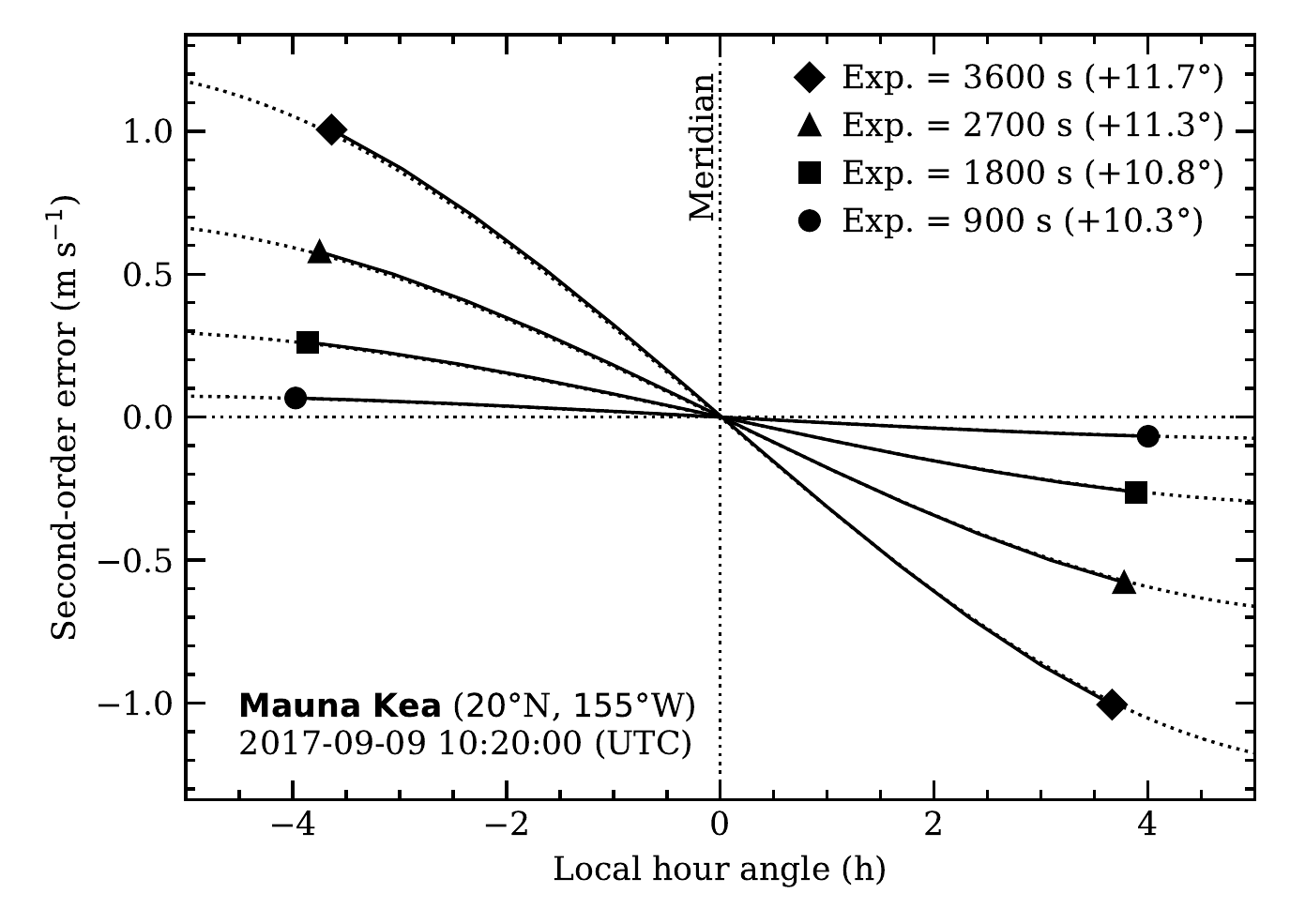}
    \caption{Second-order error, $v(\avg{t}) - \avg{v}$, simulated as function of local hour angle for various exposure times. Each point on the curves corresponds to an exposure at that hour angle. The declination and minimum/maximum hour angle is chosen for each exposure time such that the the telescope is pointing straight east/west at the geometric midpoint of the exposure, while the exposure starts/ends at $30^\circ$ altitude. The exposure meter flux is uniform.}
    \label{fig:exptime}
\end{figure}

\begin{figure}
    \centering
    \includegraphics[width=\linewidth]{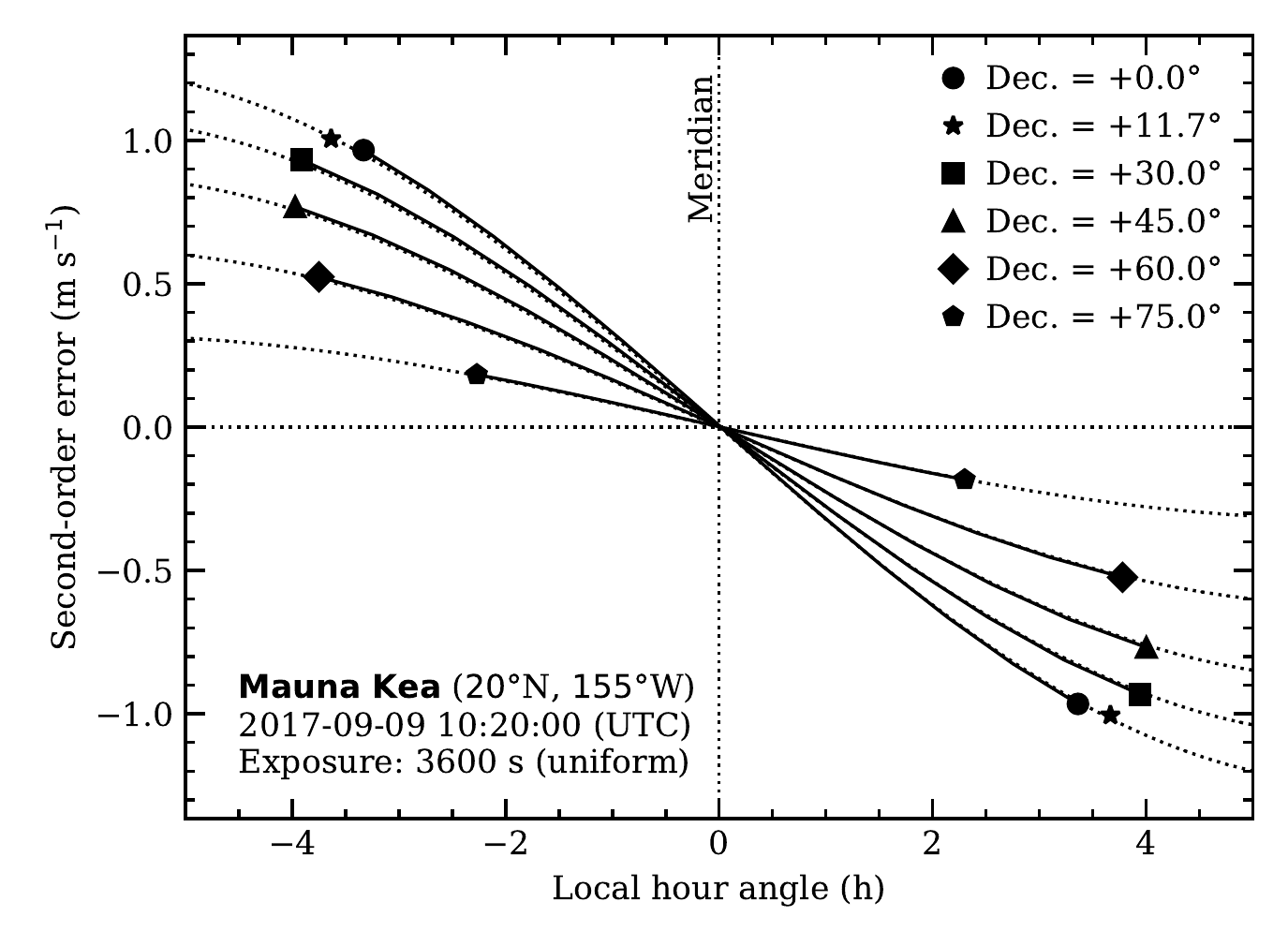}
    \caption{Second-order error, $v(\avg{t}) - \avg{v}$, simulated as function of local hour angle for various declinations. The exposure time is 1800~seconds with a uniform flux curve. All curves end at $37^\circ$ altitude, ensuring that all exposures begin and end above $30^\circ$ altitude. The two $\star$ markers indicate the maximum error for the worst declination with this altitude limit ($\delta=11.7^\circ$). The exposure meter flux is uniform.}
    \label{fig:declination}
\end{figure}

\begin{figure}
    \centering
    \includegraphics[width=\linewidth]{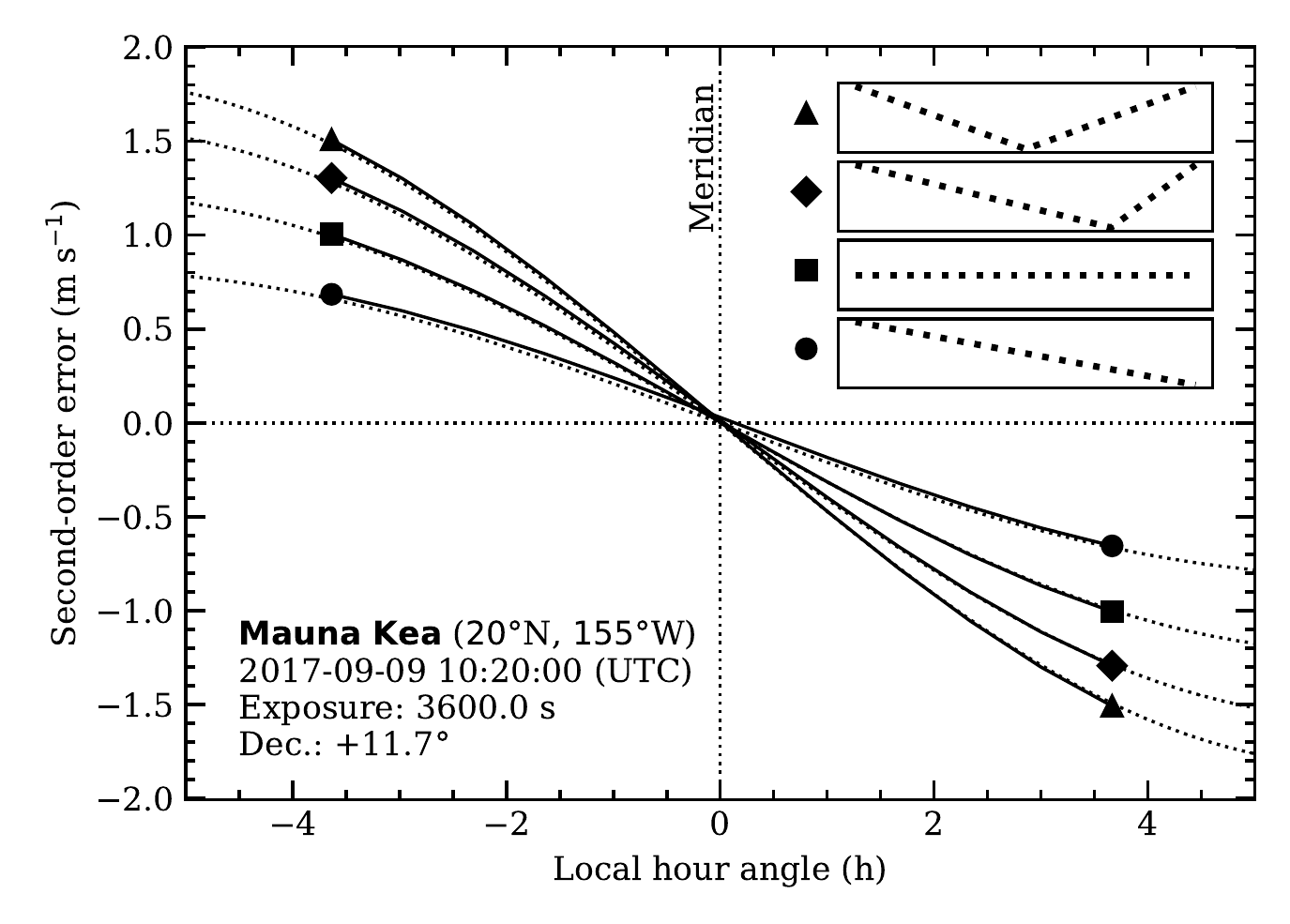}
    \caption{Second-order error, $v(\avg{t}) - \avg{v}$, simulated as function of local hour angle for various shapes of the exposure meter flux curve. The shapes are a linear ramp, a uniform exposure, a centred V-shape, and a V-shape offset from the centre. The dotted curves in the background are the analytically approximated values, made by multiplying the uniform result in Equation~\eqref{eq:err_uniform} by 0.67, 1.00, 1.29, and 1.50.}
    \label{fig:shape}
\end{figure}

In \fref{fig:exptime}-\ref{fig:shape}, we use the location of Mauna Kea Observatory ($19.8222^\circ$N, $155.4749^\circ$W, 4205~m) and select the time as 10:20 UTC (local midnight) on an arbitrarily chosen night (09-Sep 2017). For each curve, we fix the declination and vary the hour angle. We do this for various exposure times (\fref{fig:exptime}), declinations (\fref{fig:declination}), and exposure meter flux curves (\fref{fig:shape}). The solid curves show the simulated errors, and the dotted curves in the background show the analytic estimates from Equation~\eqref{eq:err_uniform}.

In \fref{fig:exptime} and \ref{fig:shape}, when the exposure starts at $30^\circ$ altitude, the declination and the minimum hour angle are chosen such that the rising target is situated at ${\rm az} = 90^\circ$ at the geometric midpoint time -- and vice versa for the maximum hour angle. In \fref{fig:declination} we simulate various declinations, while still choosing the minimum and maximum hour angle such that the exposure starts and ends at $30^\circ$ altitude or above.

The uniform 60-minute exposure at $\delta$=$+11.7^\circ$ is present in all three plots, and the size of its error ranges between  $\pm1.00\ms$ in agreement with the analytic estimate from Section~\ref{sec:derivation}. In \fref{fig:shape}, we simulate different shapes of the exposure meter flux curve, and we verify that the error increases by a factor of 1.5 when the photon flux drops and rises again during the exposure.

\begin{figure*}
    \centering
    \includegraphics[width=\textwidth]{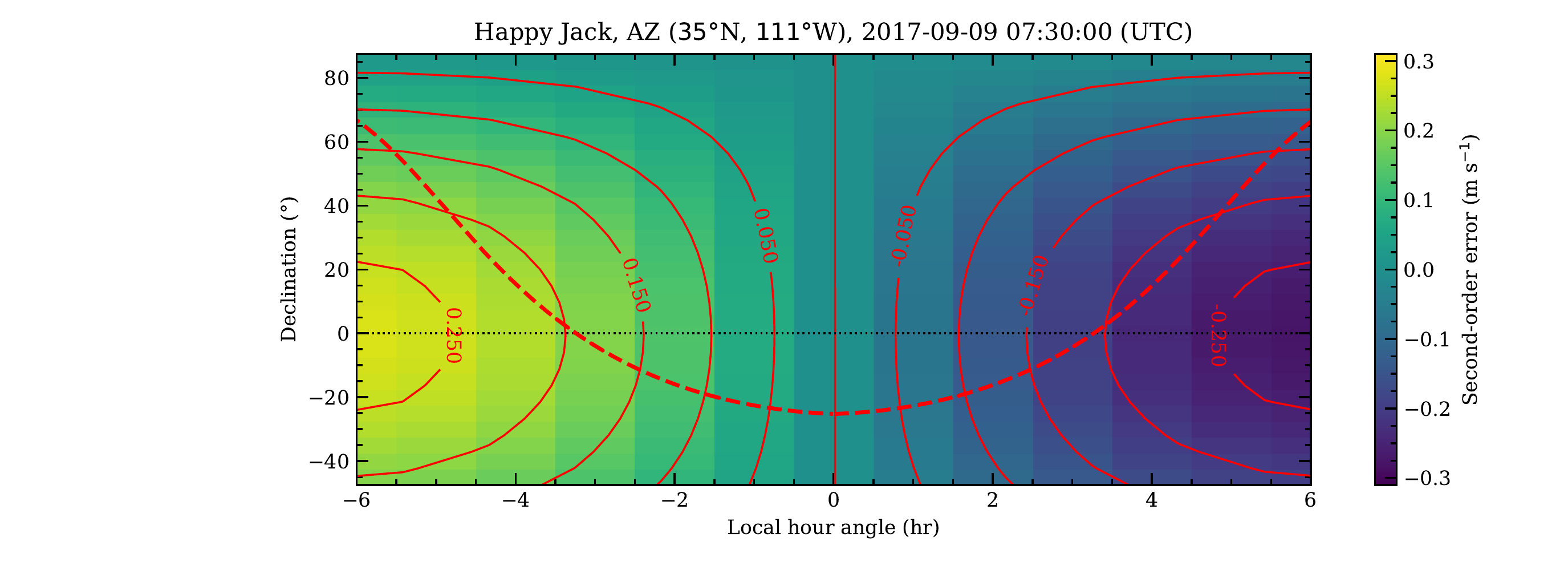}
    \includegraphics[width=\textwidth]{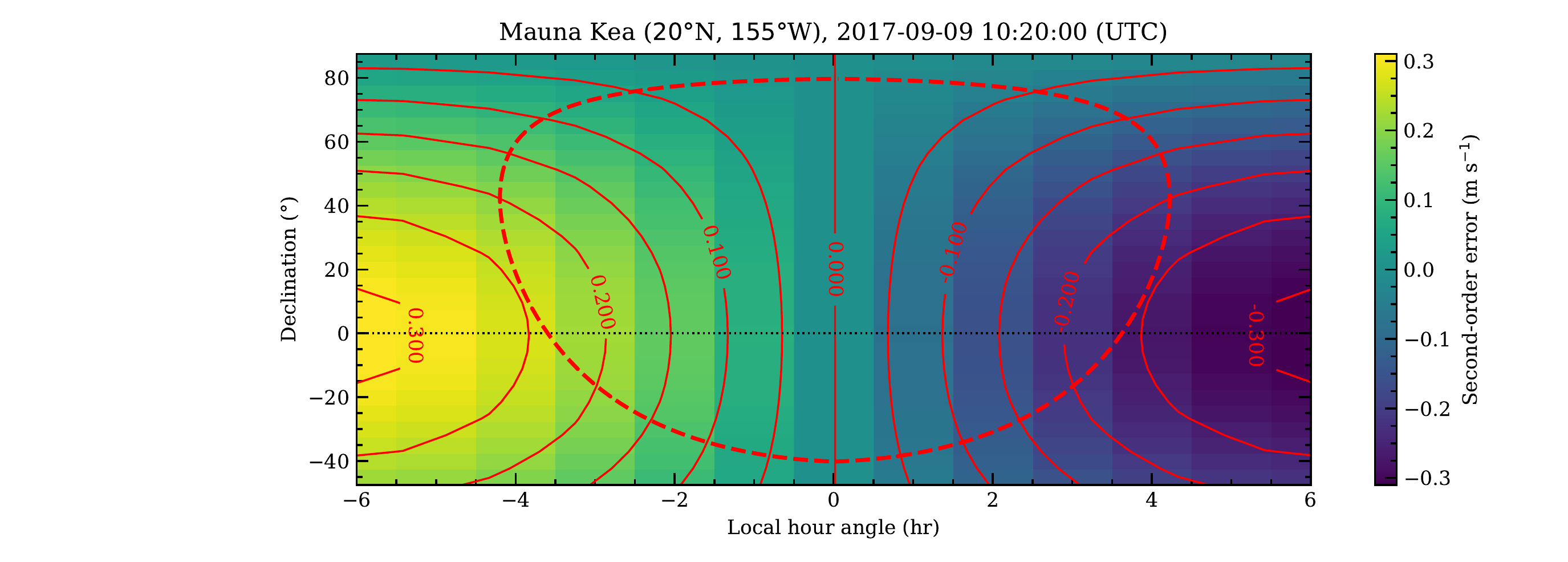}
    \includegraphics[width=\textwidth]{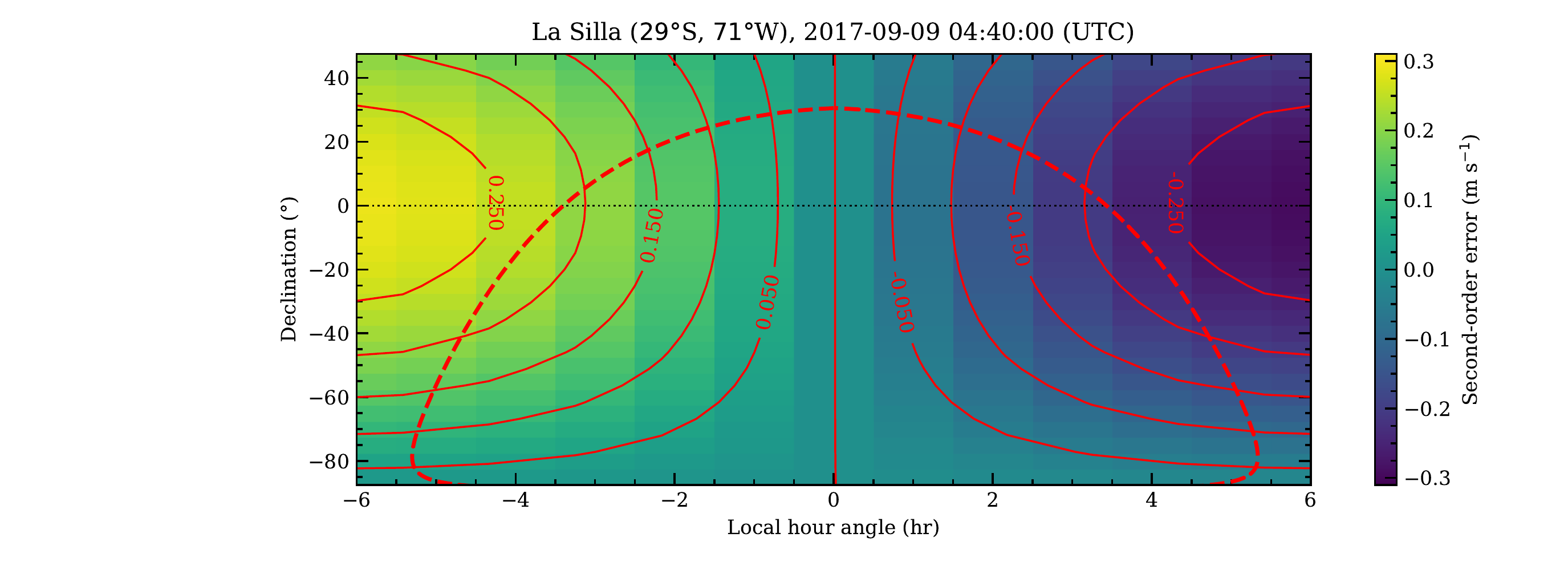}
    \caption{Second-order error, $v(\avg{t}) - \avg{v}$, simulated as function of sky position at three different observatories. We simulate a 30 minute exposure with uniform flux, observed at local midnight. The colour indicates the size of the error as a function of the target position at the geometric midpoint time; the dashed red lines indicate the altitude limit that ensures the exposure is completed above $30^\circ$, while solid red lines are contours, drawn for every 0.05\ms. The observatories are Happy Jack (EXPRES), Mauna Kea (e.g. HiRes, KPF, MAROON-X, SPIROU), and La Silla (HARPS, CORALIE); for a more comprehensive list of instruments, see e.g. \citet{Plavchan:2015}, \citet{Fischer:2016}, and \citet{Wright:2017}.}
    \label{fig:hadec}
\end{figure*}

In \fref{fig:hadec} we provide contour maps of the sky for three observatories at different latitudes, showing for a 30-minute uniform exposure how the second-order error depends on the sky position. The 30-minute curve in \fref{fig:exptime} corresponds to a horizontal cut through the central diagram in \fref{fig:hadec}. The area defined by the dashed red line contains all the observations that both start and stop at an altitude higher than $30^\circ$. The asymmetric shape of the area gives a visual understanding why the maximum effect is not reached at declination zero.

All the simulations in Figures~\ref{fig:exptime}-\ref{fig:hadec} are made with 1~Hz sampling cadence. When we generate the same plots with 0.1~Hz sampling, the results are practically identical. The way we interpret this is that our cadence is sufficiently high that the results are not affected by it. Cadence effects can be analysed in more detail by binning real exposure meter flux curves to different bin sizes. We expect the result of such an analysis to depend on the signal-to-noise level and other specifics of the particular instrument, and it is therefore beyond the scope of this manuscript.

% BC_error2 vs flux curve shape, lat=20, alt=40, Az=90

%HA-dec grid

\section{Discussion}
\label{sec:discussion}
In the preceding sections, we have demonstrated the dependence and magnitude of a second-order error in the barycentric correction when applied to longer exposures. In the following, we will discuss the possible implications for exoplanet mass measurements, and we will explore some options for mitigating errors when observing and/or analysing data, in cases where the exposure meter is broken or unavailable.

\subsection{Implications for exoplanet mass measurements}
As a new generation of PRV spectrographs are aiming at the 10\cms{} and 1\cms{} regimes, hopes are that astronomers will soon be able to detect Earth-like planets around Sun-like stars. If we were to observe the gravitational tug that our own planet exerts on the Sun, the RV semi-amplitude would be as little as 9\cms. We have shown in this paper that the second-order error on the barycentric correction can easily reach and exceed this level. 
Next-generation PRV instruments must therefore record the exposure meter flux curve for each observation, allowing the observer or an automated pipeline to calculate the photon-weighted barycentric correction. 
We recommend that the exposure meter data is stored and archived along with the observed spectrum, allowing the photon-weighted barycentric correction to be fully reproduced and checked at a later time. Although the calculation is trivial, mistakes can still happen (e.g. a broken FITS header) and the observer should always have the option to go back and recalculate the barycentric correction from scratch for each data point.
As argued by \citet{Wright:2014} and elaborated by \citet{Blackman:2017}, it is also recommended that exposure meters split the light into multiple wavelength channels in order to accommodate wavelength dependent fibre coupling and seeing, as it is done at instruments like ESPRESSO \citep{Pepe:2010,Landoni:2014} and EXPRES \citep{Jurgenson:2016,Blackman:2017}.

Over the last decade, instruments like HARPS \citep{Mayor:2003} and {\harpsn} \citep{Cosentino:2012} have demonstrated their capability to measure RVs with a precision of 1\ms{} \citep{Fischer:2016}. One could argue that the barycentric second-order error is mostly negligible at this level, but as we have seen, situations exist where it becomes a substantial fraction of or even more than 1\ms. 
The barycentric second-order errors for a set of RV measurements do not necessarily form a normal distribution -- they depend on weather, atmospheric seeing, observing strategies, and decisions (or mistakes) made by the observers. Our concern is that sometimes these errors may alter the measured velocities in ways that change the apparent amplitude of the RV signal. If that happens, it can affect the modelled mass and/or eccentricity. 

Most people seek to observe their targets close to the meridian, where the second-order effect diminishes, but sometimes scheduling constraints requires a target to be observed at higher air mass. One particular example of this is when a target is followed throughout its observing season. When it first becomes visible, in the beginning of the observing season, it will be rising in the east, and it will be observed at a large negative hour angle in order to expose at the highest possible altitude before twilight. As time passes, the target becomes visible all night and observable near the meridian. Towards the end of the season, the target is setting in the beginning of the night, and the observer needs to observe at a large positive hour angle. For a target near the celestial equator, with 30 minutes exposure time, the insufficiently corrected velocities could have an apparent $\pm 25\cms$ ``curl'' with opposite signs in the beginning and end of the observing season.

A similar situation could happen when a target instead of once per night is visited twice per night. An observer may try to separate the observations in time as much as possible by observing in the beginning and end of the night at the highest feasible air mass. This could lead to systematic barycentric correction errors where the early and late observations each are offset by e.g. $25\cms$ with opposite signs. Sometimes this type of data is fitted with a nightly offset to account for trends due to stellar activity \citep{Pepe:2013,Howard:2013}. Any RV variation within a night would thus be interpreted as Keplerian motion, and the the barycentric error could propagate to the fitted semi-amplitude.

As long as we calculate and apply the barycentric correction with proper photon weights, the RV errors due to barycentric correction will be limited to less than 1\cms{}, as in \citet{Wright:2014}, and the situations described above do not cause any particular problems with respect to barycentric correction. Otherwise, there is a risk that data from current-generation PRV instruments near 1\ms{} precision could sometimes be affected by second-order errors, and we advise caution.

\subsection{Mitigation strategies}
\noindent In cases where no flux curve is supplied with an exposure, there are still ways to mitigate the consequent errors in the barycentric correction. 

If the observer knows in advance that the exposure meter is broken or for some reason not available, it may be desirable to modify the observing strategy or even switch to other targets. We know from Equations~\eqref{eq:firstorder} and \eqref{eq:taylor_weighted} that targets near the celestial equator are more strongly affected by both errors, so targets at high declination would be better suited for a night like that. We also know that the size of the first-order error can be reduced by shortening the exposure time or observing at larger local hour angle; the latter however comes at the price of higher air mass.  In order to keep the flux as uniform as possible, observing through clouds without an exposure meter is not advisable.

\begin{figure}
    \centering
    \includegraphics[width=\linewidth]{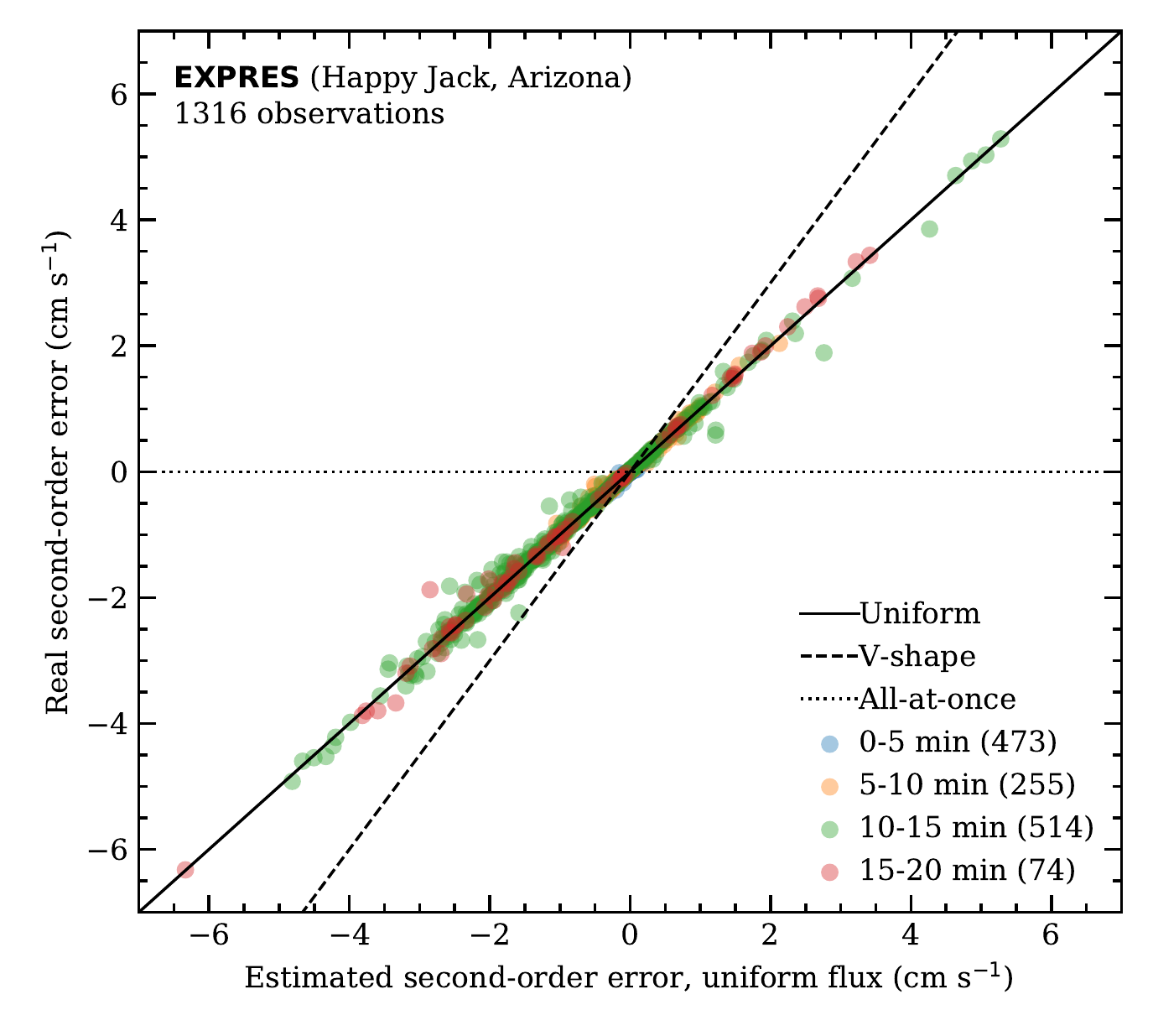} 
    \caption{For nearly one year of EXPRES observations, we compare the actual second-order error of each measurement with the uniform flux approximation. The multi-channel exposure meter data has been binned to one channel. EXPRES mainly observes bright targets, so the exposure times are short ($\leq 1200$ seconds), resulting in errors of only a few\,\cms{}. Nevertheless, the data shows that the uniform flux curve is a good approximation if we for some reason do not have access to the exposure meter flux curve when we want to calculate the second-order error.}
    \label{fig:expres_uniform}
\end{figure}

Observers may also find themselves at an instrument where the exposure meter only delivers the photon-weighted midpoint time, $\avg{t}$, and does not preserve the flux curve. In this case, the first-order effect is accounted for by calculating the barycentric correction at $\avg{t}$, and our only concern is the second-order effect. Since it is proportional to $\sin (\psi)$, it can be easily minimised by observing near to the meridian. Decreasing the exposure time or choosing bright targets with shorter exposure times will also help. \fref{fig:hadec} gives a visual idea about what areas of the local sky are within a given error margin at three different observatory sites.

When dealing with PRV data that supplies the photon-weighted midpoint time and yet no flux curve, the simplest assumption is that the flux was uniformly distributed in time\footnote{One may (rightfully) argue that a uniform flux distribution over the entire exposure time entails that $\avg{t}=t_0$. For a more stringent approach, assume a step function set to zero in the beginning or the end of the exposure, such that the photon-weighted midpoint time $\avg{t}$ is reproduced and the exposure time $\Delta t$ is effectively shortened by $2 |\avg{t}-t_0|$.}. This allows us to add a second-order correction to the calculated barycentric correction, $v(\avg{t})$. To obtain $\avg{v}$, we can either subtract the result of Equation~\eqref{eq:err_uniform} or get the difference from a simple numerical simulation as in Section~\ref{sec:simulations}. In fact, if we do not apply a second-order correction to $v(\avg{t})$, we implicitly assume that all photons arrived in the same instant at the photon-weighted midpoint time. We argue that the assumption of a uniform flux distribution is much more realistic. \fref{fig:shape} shows how the second-order error of three different flux scenarios are all closer to the uniform scenario than zero. 

In \fref{fig:expres_uniform} we calculate the second-order error using real exposure meter data from 1315 observations with EXPRES \citep[see][]{Blackman:2019}, combining all colour channels to one. The data was obtained during commissioning between March 2018 and February 2019. No exposures were longer than 20 minutes, and as a result the second-order errors are relatively small. We plot the real second-order errors against the errors estimated with Equation~\eqref{eq:err_uniform} for a uniform flux curve. The plot shows that within 1\cms, the uniform flux approximation is indeed a realistic assumption, and much more realistic than setting the error to zero. There is no indication that longer exposure times would change this picture.

In the rare case of dealing with data that completely lacks information about the temporal flux distribution, there is no way of knowing if the individual observations have large first-order barycentric errors. One can in principle use Equation~\eqref{eq:firstorder} to estimate the RV error bars, but it requires a qualified guess at the offset $(\avg{t}-t_0)$. In our experience, a typical offset is of order 5\% of the exposure time, with some larger, negative offsets from when the star was lost on the fibre before the end of the exposure. The actual distribution depends on the instrument and the local sky conditions.

\section{Conclusions}
%The last numbered section should briefly summarise what has been done, and describe
%the final conclusions which the authors draw from their work.
\label{sec:conclusions}
This paper describes how to correctly apply a barycentric correction to an exposure using the exposure meter flux curve or an equivalent measure. Rather than following the common practice of computing a single value at the photon-weighted midpoint time, we argue why the correct approach is to apply the photon weights to the barycentric correction itself. The systematic difference between the two methods can be described as a second-order effect that leads to systematic RV errors of up to 0.25\ms{} for a 30 minute exposure and 1.0\ms{} for a 60 minute exposure. Certain observing scenarios, when the photon flux drops and rises again during an exposure, can amplify the error by a factor of $\sim$1.5.

Considering that next-generation PRV spectrographs are aiming for the 10\cms{} regime and beyond, it will be absolutely necessary to account for this effect. Instruments of less precision (1\ms) are likewise affected for some observations, depending on where on the sky and the exposure time. For instruments that already measure the photon-weighted midpoint, it should be somewhat trivial to compute the photon-weighted barycentric correction in order to eliminate unnecessary systematic errors. We also show that it is possible to partially mitigate second-order errors even in archive data that was stored without the flux curve.

\section*{Acknowledgements}
%The Acknowledgements section is not numbered. Here you can thank helpful
%colleagues, acknowledge funding agencies, telescopes and facilities used etc.
%Try to keep it short.
We would like to thank the referee, Artie Hatzes, for carefully reading the manuscript and supporting its conclusions.
We also thank the members of the EXPRES and {\harpsn} GTO teams for useful input and fruitful discussions. 
Our simulations involved an extensive amount of coordinate transformations, made possible with the excellent \texttt{astropy.coordinates} module\footnote{http://astropy.org} \citep{astropy:2013, astropy:2018}. All figures were prepared with Matplotlib\footnote{http://matplotlib.org} \citep{Hunter:2007}.
The Center for Exoplanets and Habitable Worlds is supported by the Pennsylvania State University, the Eberly College of Science, and the Pennsylvania Space Grant Consortium.

%%%%%%%%%%%%%%%%%%%%%%%%%%%%%%%%%%%%%%%%%%%%%%%%%%

%%%%%%%%%%%%%%%%%%%% REFERENCES %%%%%%%%%%%%%%%%%%

% The best way to enter references is to use BibTeX:

\bibliographystyle{mnras}
\bibliography{references} % if your bibtex file is called example.bib

%%%%%%%%%%%%%%%%%%%%%%%%%%%%%%%%%%%%%%%%%%%%%%%%%%

%%%%%%%%%%%%%%%%% APPENDICES %%%%%%%%%%%%%%%%%%%%%

%\appendix

%\section{Some extra material}

%If you want to present additional material which would interrupt the flow of the main paper,
%it can be placed in an Appendix which appears after the list of references.

%%%%%%%%%%%%%%%%%%%%%%%%%%%%%%%%%%%%%%%%%%%%%%%%%%

% Don't change these lines
\bsp	% typesetting comment
\label{lastpage}
\end{document}